\documentclass[12pt]{article}
\begin{document}

\begin{center} {\bf\LARGE Why Poincare symmetry is a good approximate symmetry in particle theory}\end{center}
\begin{center}Felix M Lev\end{center}
\begin{center} Independent Researcher, San Diego, CA USA\end{center}
\begin{center}Email: felixlev314@gmail.com\end{center}
\begin{abstract}
As shown by  Dyson in his famous paper "Missed Opportunities", it follows even from purely mathematical considerations that quantum Poincare symmetry is a special degenerate case of quantum de Sitter symmetries. Then the usual explanation of why  in particle physics Poincare symmetry works with a very high accuracy is as follows. A theory in de Sitter space becomes a theory in Minkowski space when the radius of de Sitter space is very high. However, the answer to this question must be given only in terms of quantum concepts while 
de Sitter and Minkowski spaces are purely classical concepts.  Quantum Poincare symmetry is a good approximate symmetry if the eigenvalues of the representation operators
$M_{4\mu}$ 
of the anti-de Sitter algebra are much greater than the eigenvalues of the operators $M_{\mu\nu}$ ($\mu,\nu=0,1,2,3$).
We explicitly show that this is the case in the Flato-Fronsdal approach where elementary particles in the standard theory are bound states of two Dirac singletons.
\end{abstract}

\begin{flushleft}Keywords: irreducible representations; de Sitter supersymmetry;
Dirac supersingletons; accuracy of Poincare symmetry \end{flushleft}

\section{Problem statement}
\label{statement}

In quantum field theory (QFT), relativistic (Poincare) symmetry is explained as follows.
Poincare group is the group of motions of Minkowski space and the quantum system under consideration (which, in the general case, can consist of an arbitrary number of interacting elementary particles) should be described by unitary representations of this group. This implies
 (see e.g., Sec. 1.3 in the textbook \cite{Novozhilov}),  that the representation generators should commute according to the commutation relations of the Poincare Lie algebra:
\begin{eqnarray}
&[P^{\mu},P^{\nu}]=0,\quad [P^{\mu},M^{\nu\rho}]=-i(\eta^{\mu\rho}P^{\nu}-
\eta^{\mu\nu}P^{\rho}),\nonumber\\
&[M^{\mu\nu},M^{\rho\sigma}]=-i (\eta^{\mu\rho}M^{\nu\sigma}+\eta^{\nu\sigma}M^{\mu\rho}-
\eta^{\mu\sigma}M^{\nu\rho}-\eta^{\nu\rho}M^{\mu\sigma})
\label{PCR}
\end{eqnarray}
where $\mu,\nu=0,1,2,3$, $\eta^{\mu\nu}=0$ if $\mu\neq \nu$, $\eta^{00}=-\eta^{11}=
-\eta^{22}=-\eta^{33}=1$,
$P^{\mu}$ are the four-momentum operators and  $M^{\mu\nu}$ are the Lorentz angular momentum operators. This is in the spirit of 
the Erlangen Program proposed by Felix Klein in 1872.
However, 
the description (\ref{PCR}) does not involve Poincare group and Minkowski space at all.

As noted in \cite{book}, in quantum theory, background space is only a mathematical concept because each physical quantity should be described by an operator while there are no operators for the coordinates of background space. This space is not used in relativistic quantum theory for describing irreducible representations (IRs) for elementary particles. According to the Heisenberg $S$-matrix program,  transformations from the
Poincare group are not used because it is possible to describe only transitions of states from the infinite past when $t\to -\infty$ to the distant future 
when $t\to +\infty$.  Here, systems should be described only by observable physical quantities --- momenta and angular momenta. So, {\it symmetry at the quantum level should be defined
by commutation relations of the symmetry algebra rather
than by a background space and its group of motions}  (see e.g., \cite{book} for more details). In particular,
Eqs. (\ref{PCR}) should be treated {\it as the definition of Poincare symmetry at the
quantum level.} 

As noted by Dyson in his famous paper "Missed Opportunities" \cite{Dyson}: 
\begin{itemize}
\item a) Quantum Poincare
theory is more general than Galilei one: 
the latter can be obtained from the former by contraction $c\to\infty$. 
\item b) de Sitter (dS) and anti-de Sitter (AdS) 
quantum theories are more general than Poincare one:
the latter can be obtained from the former by contraction $R\to\infty$
where $R$ is a parameter with the dimension $length$. 
\item c) At the same time, being semisimple, dS and AdS groups cannot be obtained from more symmetric ones by contraction.  
\end{itemize}

As noted above, quantum symmetry should be defined in terms of Lie algebras, 
and in \cite{book}, the
statements a)-c) have been reformulated in such terms. In addition, quantum theory is more general than classical one 
because the classical symmetry algebra can be obtained from the quantum one by contraction $\hbar\to 0$. As a consequence, the most general description in terms of ten-dimensional Lie algebras should be defined in terms of quantum dS or AdS symmetry. 

The definition of these symmetries is described in an extensive literature 
(see e.g., \cite{book,Evans}): the angular momentum operators  
$M^{ab}$ ($a,b=0,1,2,3,4$, $M^{ab}=-M^{ba}$) should satisfy the
commutation relations:
\begin{equation}
[M^{ab},M^{cd}]=-i (\eta^{ac}M^{bd}+\eta^{bd}M^{ac}-
\eta^{ad}M^{bc}-\eta^{bc}M^{ad})
\label{CR}
\end{equation}
Here the tensor $\eta^{ab}$ is such that $\eta^{ab}=\eta_{ab}$, $\eta^{ab}=0$ if $a\neq b$, 
$\eta^{00}=-\eta^{11}=-\eta^{22}=-\eta^{33}=1$, 
$\eta^{44}=\mp 1$ for the dS and AdS symmetries, respectively, and this tensor is used 
to raise and lower the indices of the operators $M^{ab}$. Eqs. (\ref{CR}) demonstrate that quantum dS and AdS theories do not involve the dimensional parameters $(c,\hbar,R)$, and this is
a consequence of the fact that $(kg,m,s)$ are meaningful only at the macroscopic level.  
These expressions {\it define dS and AdS symmetries at the quantum level} and they do not involve dS and AdS groups and spaces \cite{ book}. 

The contraction from dS or AdS symmetry to Poincare one is defined as follows: if the momentum operators $P^{\mu}$
{\it are defined}  as $P^{\mu}=M^{4\mu}/R$ ($\mu=0,1,2,3$) and when $R\to\infty$ $M^{4\mu}\to\infty$ but the quantities
$P^{\mu}$ are finite, then Eqs. (\ref{CR}) become Eqs. (\ref{PCR}). 
As a consequence, as shown in Sec. 1.3 of \cite{book}, dS and AdS symmetries are more general (fundamental) than Poincare symmetry. Note that $R$ has nothing to do with the radius of dS or AdS spaces. 

In the literature, this issue is discussed with numerous examples, but, as shown in Sec. 1.3 of \cite{book}, with any desired accuracy, any result of Poincare symmetry can be reproduced in dS or AdS symmetries at some choice of $R$ but when the limit $R\to\infty$ has already been taken, Poincare symmetry cannot reproduce those results
of dS and AdS symmetries where it is important that $R$ is finite and not infinitely large. 

There is an analogy here with the fact that, since Galilei algebra can be obtained from Poincare one by contraction, Poincare symmetry is more general (fundamental) than Galilei one. Namely, it can be shown \cite{book} that any result of Galilei symmetry can be reproduced in Poincare symmetry at some choice of $c$ but when the limit $c\to\infty$ has already been taken, Galilei symmetry cannot reproduce those results of Poincare symmetry where it is important that $c$ is finite and not infinitely large.

At the classical (non-quantum) level, the transition from dS or AdS symmetry to Poincare one is explained as follows. When the radius of dS or AdS space becomes infinitely large, the angular momentum $M$ of a particle moving in this space also becomes infinitely large. When dS or AdS space transforms into flat Minkowski space, the motion of a particle in such space must be described by the momentum $p=M/R$ which is finite in this limit.

One can raise a question why Poincare symmetry works with great accuracy in particle physics.
At the classical level, the explanation is that we live in dS or AdS space whose radius 
is very large. The cosmological data show that this is indeed the case because at the present stage of the universe, this radius is of the order of $10^{26}m$ \cite{Lambda}. However, as noted above, the concept of background space is purely macroscopic that should not be used in particle theory. Therefore, a question arises whether the answer can be given within the framework of purely quantum theory, without involving classical concepts. However, there is no such explanation in the literature at a purely quantum level.

As follows from the above definition of contraction from dS or AdS algebra to Poincare one, Poincare symmetry works with a high accuracy, provided that such states play the major role in which the eigenvalues
of the operators $M^{4\mu}$ are much greater than the eigenvalues of the operators
$M^{\mu\nu}$ ($\mu,\nu$=0,1,2,3). In this paper, we propose a scenario that describes such a situation.

The paper is organized as follows. In Sec. \ref{supersymmetry} we explain why
supersymmetric AdS symmetry is more general (fundamental) than standard AdS
symmetry. In Sec. \ref{CPT} we describe how the CPT transformation works at the
quantum level. Then in Sec. \ref{Scenario}
it is explicitly shown that there exist scenarios when Poincare symmetry works with a high accuracy.

\section{Supersymmetry}
\label{supersymmetry}

Since dS and AdS symmetries are more general than Poincare symmetry (see the preceding  section), it is natural to consider supersymmetric generalizations of dS and AdS symmetries. Such generalizations exist
in the AdS case but do not exist in the dS one. As shown in \cite{Lambda}, in standard quantum theory, dS symmetry is more general than AdS one, and it may be a reason why supersymmetry has not been discovered yet. However, standard quantum theory is a special degenerate case of a quantum theory
over a finite ring  of characteristic $p$ (FQT) in a formal limit $p\to\infty$ \cite{book}, and in FQT, dS and AdS symmetries are equivalent. 
For this reason, in what follows we will consider supersymmetric generalizations of AdS symmetry.

By analogy with representations of the Poincare superalgebra,  
representations of the osp(1,4) superalgebra also are described by 14 operators: ten operators of the so(2,3) algebra commute with each other as
in Eqs. (\ref{CR}), anticommutators of the four fermionic
operators are linear combinations of the so(2,3) operators and
commutators of the fermionic operators with the so(2,3) operators are
linear combinations of the former. However, a fundamental fact of the os(1,4) supersymmetry is that 
the osp(1,4) superalgebra can be described exclusively in terms
of the fermionic operators because the anticommutators of four operators form ten
independent linear combinations. Therefore, ten bosonic
operators can be expressed in terms of fermionic ones. This implies that (by analogy with the treatment of the Dirac equation as a square root from the Klein-Gordon equation) the
osp(1,4) symmetry is an implementation of the idea that
supersymmetry is the extraction of the square root from the
usual symmetry .

The fermionic operators $(d_1',d_2',d_1'',d_2'')$ of the osp(1,4) superalgebra satisfy the following
relations. If $(A,B,C)$ are any fermionic operators, [...,...]
is used to denote a commutator and $\{...,...\}$ to denote an
anticommutator then
\begin{equation}
[A,\{ B,C\} ]=F(A,B)C + F(A,C)B
\label{S30}
\end{equation}
where the form $F(A,B)$ is skew symmetric, $F(d_j',d_j")=1$
$(j=1,2)$ and the other independent values of $F(A,B)$ are
equal to zero. 

As shown in \cite{book,Heidenreich}, the operators $M^{ab}$ in 
Eqs. (\ref{CR}) can be expressed through bilinear combinations of the fermionic operators:
\begin{eqnarray}
&&h_1=\{d_1',d_1''\},\,\,h_2=\{d_2',d_2''\},\,\,M_{04}=h_1+h_2,\,\,
M_{12}=L_z=h_1-h_2\nonumber\\
&&L_+=\{d_2',d_1''\},\,\, L_-=\{d_1',d_2''\},\,\, M_{23}=L_x=L_++L_-\nonumber\\
&&M_{31}=L_y=
-i(L_+-L_-),\,\, M_{14}=(d_2'')^2+(d_2')^2-(d_1'')^2-(d_1')^2\nonumber\\
&&M_{24}=i[(d_1'')^2+(d_2'')^2-(d_1')^2-(d_2')^2]\nonumber\\
&& M_{34}=\{d_1',d_2'\}+\{d_1'',d_2''\},\,\,M_{30}=-i[\{d_1'',d_2''\}-\{d_1',d_2'\}]\nonumber\\
&&M_{10}=i[(d_1'')^2-(d_1')^2-(d_2'')^2+(d_2')^2]\nonumber\\
&&M_{20}=(d_1'')^2+(d_2'')^2+(d_1')^2+(d_2')^2
\label{Mab}
\end{eqnarray}
where ${\bf L}=(L_x,L_y,L_z)$ is the standard operator of three-dimensional rotations.

For finding IRs, we require
the existence of the vector $e_0$ such that:
\begin{eqnarray}
d_j'e_0=d_2'd_1''e_0=0, \quad d_j'd_j''e_0=q_je_0\quad (j=1,2)
\label{S32A}
\end{eqnarray}
\begin{sloppypar}
These conditions show that the Cartan subalgebra operators are $\{d_j',d_j''\}\,\,(j=1,2)$. 
The full representation space can be obtained by successively
acting by the operators $d_j',d_j''$ on $e_0$ and taking all
possible linear combinations of such vectors. The theory of IRs of the
osp(1,4) algebra has been developed by Heidenreich \cite{Heidenreich},
and in \cite{book} this theory has been generalized to the case of FQT.
 \end{sloppypar}
\section{CPT transformation in osp(1,4) invariant theory}
\label{CPT}

In Poincare invariant particle theory, the CPT transformation is considered the most general discrete spacetime transformation. Based on what was said in Sec. \ref{statement}, at the quantum level, this transformation should be considered not from the point of view of Minkowski space, but at the operator level. We use $\theta$ to denote the operator corresponding to the quantum CPT transformation. As Wigner noted \cite{Wigner}, since the sign of the energy must remain positive under the $\theta$ transformation, the operator $\theta$ must be not unitary, but antiunitary, that is, it can be represented as $\theta=\beta K$ where $\beta$ is a unitary operator, and $K$ is the complex conjugation operator. As shown by Schwinger
\cite{Schwinger}, the problem of the sign of energy can also be solved if instead of the antiunitary transformation the transpose operation is used. In this paper we use Wigner's approach.

As shown in \cite{Wigner} (see also \cite{Novozhilov}), the operator $\theta$ transforms the operators in Eq. (\ref{PCR}) as
\begin{equation}
\theta P_{\mu} \theta^{-1}=P_{\mu}, \quad \theta M_{\mu\nu} \theta^{-1}=-M_{\mu\nu}
\label{Poincaretheta}
\end{equation}
A question arises of how to generalize these relationships to the case of dS and AdS theories
and, as noted in Sec. \ref{statement}, this generalization should not involve dS and AdS spaces. The issue of CPT transformation in such theories has been considered by many authors. However, to the best of our knowledge, these authors considered the CPT transformation only from the point of view of transformations of fields on the dS and AdS spaces and did not consider a direct generalization of Eq. (\ref{Poincaretheta}). Moreover, as noted in Sec. \ref{supersymmetry}, the superalgebra osp(1,4) is a generalization of the algebra so(2,3) to the case of supersymmetry, and Eqs. (\ref{Poincaretheta}) have not been generalized to representations of this superalgebra.

For this purpose, it is necessary to define how the operator $\theta$ transforms the operators
$(d_1',d_2',d_1'',d_2'')$. We define such a transformation as follows:
\begin{equation}
\theta d_1' \theta^{-1}=-id_2',\,\,\theta d_2' \theta^{-1}=id_1',\,\, 
\theta d_1'' \theta^{-1}=id_2'',\,\,\theta d_2'' \theta^{-1}=-id_1'' 
\label{osptheta}
\end{equation}
It is easy to see that the second of these relations follows from the first, and the fourth follows from the third, because the operator $\theta$ is antiunitary.

Now, based on Eqs. (\ref{Mab}) and (\ref{osptheta}) we conclude that
\begin{equation}
\theta M_{4\mu} \theta^{-1}=M_{4\mu}, \quad \theta M_{\mu\nu} \theta^{-1}=-M_{\mu\nu},\quad \mu,\nu =0,1,2,3
\label{AdStheta}
\end{equation}
and this is a generalization of Eq. (\ref{Poincaretheta}) to the case of 
representations of the algebra so(2,3) because, as noted in Sec. \ref{statement}, when contracting representations of the algebra AdS into representations of the Poincare algebra, the operators $M_{\mu\nu}$ are not affected, and the operators $M_{4\mu}$ go into $P_{\mu}$.
This result is also natural from the observation that, as it is easy to see, Eqs. (\ref{CR}) are invariant under substitutions
$$M_{4\mu}\to M_{4\mu}, \quad M_{\mu\nu}\to -M_{\mu\nu},\quad i\to -i$$
That these substitutions involve $i\to -i$ follows from the fact that the operator $\theta$ is 
antiunitary.

\section{Why Poincare symmetry in particle theory works with high accuracy}
\label{Scenario}

As shown in the seminal paper by Flato and Fronsdal \cite{FF}
(see also \cite{HeidenreichS}), each massless IR in standard AdS theory can be 
constructed from the tensor product of two singleton IRs discovered by Dirac in his 
famous paper \cite{DiracS} titled "A Remarkable Representation of the 3 + 2 de Sitter group". In view of this result, 
various authors gave arguments that only Dirac singletons can be true elementary particles. 
For the first time, this idea was discussed, apparently, in \cite{FFS}, and in \cite{book,Bekaert1,Ponomarev1,Ponomarev2,Bekaert2,supergrav} it was discussed from the point of view of quantum theory over finite mathematics, AdS/CFT correspondence and supergravity. In this paper we will present only one of these arguments.

If $m$ is the mass of a particle in relativistic quantum theory and $\mu$ is the
mass of this particle in AdS quantum theory then, as follows from the definition of contraction from the AdS algebra to the Poincare algebra, $\mu=mR$. 
As explained in \cite{Lambda}, $R$ has nothing to do with the radius of classical AdS
space, $R$ is fundamental to the same extent
as $c$ and $\hbar$, and the problem why the value of $R$ is as is does not arise.
As already noted, at the present stage of the universe,
$R$ is of the order of $10^{26}m$. Therefore, even for elementary particles, the  AdS masses are very large. For example, the AdS mass of the electron is of the order of $10^{39}$ and this might be an indication that the electron is not a true elementary particle. 

As noted in the literature, in standard theory there are four types of singletons: Di, Rac and their antiparticles. In the supersymmetric theory, Di and Rac are combined into a supersingleton and therefore in this theory there are only two types of singletons: the supersingleton and its antiparticle. However, as shown in \cite{book}, in FQT a particle and its antiparticle are combined into one object and therefore in FQT only the supersingleton remains.

The IR describing the supersingleton is constructed as follows: in Eq. (\ref{S32A}), $q_1$ and $q_2$ are chosen equal $q_0=1/2$ in standard theory
over complex numbers and $q_0=(p+1)/2$ in FQT, where $p$ is the characteristic of the ring
and in the latter case, $p$ is odd.

As shown in \cite{book}, the operators 
$d_1''$ and $d_2''$ commute in the space of the supersingleton IR. The basis of this IR can be chosen as $e(j,k)=(d_1'')^j(d_2'')^ke_0$ where $j,k=0,1,...\infty$ in   
standard theory and $j,k=0,1,...p-1$ in FQT. Then it can be shown \cite{book} that
\begin{equation}
d_1'e(j,k)=\frac{1}{2}je(j-1,k),\,\,d_2'e(j,k)=\frac{1}{2}ke(j,k-1)
\label{d'}
\end{equation} 
in standard theory, and $1/2$ should be replaced by $(p+1)/2$ in FQT. 

Now we can consider the problem posed in Sec. \ref{statement}: why in particle theory, 
the eigenvalues of the operators $M_{4\mu}$ are much greater than the eigenvalues of the operators $M_{\mu\nu}$ ($\mu,\nu$=0,1,2,3). As noted in Sec. \ref{statement}, this problem must be solved exclusively within the framework of quantum theory, without involving such classical concepts as Minkowski, dS or AdS spaces.

We consider the case when a particle that is treated as elementary in the standard theory is described by the tensor product of two singletons. As explained in \cite{book}, in standard theory it is believed that such singletons do not interact with each other, but in FQT they actually interact. The representation operators for the two-singleton system are the sums of the corresponding single-singleton operators: $M_{ab}=M_{ab}^{(1)}+M_{ab}^{(2)}$. This means that if $\Psi_1$ is the state of supersingleton 1, and $\Psi_2$ is the state of supersingleton 2 then the operator $M_{ab}$ acts on the tensor product of these supersingletons as 
\begin{equation}
M_{ab}(\Psi_1\times \Psi_2)=(M_{ab}^{(1)}\Psi_1\times \Psi_2)+
(\Psi_1\times M_{ab}^{(2)} \Psi_2)
\label{tensorproduct}
\end{equation}

Let us first consider the case of neutral particles, which are considered elementary in the standard theory. They can be treated as singleton-antisingleton bound states. 
Let singleton 1 be considered a particle for which the AdS algebra representations are described by the operators (\ref{CR}). The question arises what representations of the AdS algebra should describe singleton 2, which is interpreted as the antiparticle for  singleton 1. In standard theory, the transition particle$\rightarrow$antiparticle can be made in several ways, for example, by transformations $C$, $CP$ 
and $CPT$. Since $C$ and $CP$ symmetries are not exact symmetries of nature, there are known cases when the operators $C$ and $CP$, acting on physical states, give states that do not exist in nature. However, since $CPT$ symmetry is considered exact, the $CPT$ transformation applied to some state will necessarily give another state that exists in nature. As noted in Sec. \ref{CPT}, at the quantum level the $CPT$ the transformation is described by the operator $\theta$ which converts the representation operators according to Eqs. (\ref{AdStheta}). 

Therefore, it is natural to assume that $M_{ab}^{(2)}=\theta M_{ab}^{(1)}\theta^{-1}$.
Then, if $\Psi_2=\theta \Psi_1$ and for some value of $\mu$, 
$M_{4\mu}^{(1)}\Psi_1=\lambda_1 \Psi_1$ then, as follows from Eqs. (\ref{AdStheta}), $M_{4\mu}^{(2)}\Psi_2=\lambda_1 \Psi_2$ because $\lambda_1$ is real. Therefore, as follows from Eq. (\ref{tensorproduct}) 
\begin{equation}
	M_{4\mu}(\Psi_1\times \Psi_2)=2\lambda_1 (\Psi_1\times \Psi_2)
	\label{2lambda}
\end{equation}
At the same time, if for some values of $\mu$ and $\nu$, 
$M_{\mu\nu}^{(1)}\Psi_1=\lambda_2 \Psi_1$ then, as follows from Eqs. (\ref{AdStheta}), $M_{\mu\nu}^{(2)}\Psi_2=-\lambda_2 \Psi_2$
and therefore, as follows from Eq. (\ref{tensorproduct}) 
\begin{equation}
M_{\mu\nu}(\Psi_1\times \Psi_2)=0
\label{0}
\end{equation}

Now we have a natural explanation of the fact that, for a system consisting of supersingleton and antisupersingleton in a bound state, the eigenvalues of the operators
$M_{4\mu}$ are much greater than the eigenvalues of the operators $M_{\mu\nu}$:
as follows from Eqs. (\ref{AdStheta},\ref{2lambda},\ref{0}), the eigenvalues of operators $M_{4\mu}$ for individual supersingletons are included in the two-particle operators $M_{4\mu}$ with the same signs, while the eigenvalues of the operators $M_{\mu\nu}$ - with different ones. Therefore, we have a natural explanation of the fact that for a particle consisting  
from supersingleton and antisupersingleton, Poincare symmetry works with high accuracy.

Let us now consider the case when supersingletons entering a bound state are not antiparticles for each other. Then the IR of the AdS algebra for each supersingleton in a bound state is described by the operators satisfying the relations (\ref{CR}) and there is no need to use the operator $\theta$. Even when we work in FQT and consider states of supersingletons in which the quantum numbers $j,k$ are much less than $p$, then, with great accuracy, we can apply standard mathematics. We assume that, although the numbers $j,k$ can be very large, they are still much less than  $p$. Therefore, in what follows,  we consider Dirac supersingletons only within the framework of standard mathematics.

We now treat $(d_1',d_2',d_1'',d_2'')$ as the operators in the Hilbert space 
related by Hermitian conjugation as $(d_1')^*=d_1''$ and $(d_2')^*=d_2''$. Then, as follows
from Eqs. (\ref{d'}), the norm squared of $e(j,k)$ equals
\begin{equation}
\label{norm}
||e(j,k)||^2=\frac{j!k!}{2^{j+k}}
\end{equation}
and the normalized basis vectors can be defined asn
\begin{equation}
\label{normalized}
{\tilde e}(j,k)=(\frac{2^{j+k}}{j!k!})^{1/2}e(j,k)
\end{equation}

In particle scattering experiments, the four-momenta, angular momenta and spatial coordinates of initial and final particles are known with great accuracy. Therefore, for each of those quantities, quantum mechanical uncertainties are much less than mean values. As shown in \cite{book}, for free
particles, each operator $M_{ab}$ can be expressed in terms of the four-momenta, angular momenta and position operators. Therefore, for each operator $M_{ab}$, quantum mechanical uncertainties are also much less than mean values. In the literature on quantum mechanics 
(see e.g., \cite{Bohm}), this situation is characterized such that 
"The classical limit or correspondence limit is the ability of a physical theory to approximate or "recover" classical mechanics when considered over special values of its parameters." For the motion of a particle, this means that its de Broglie wavelength changes little over distances of the order of the size of this particle. In \cite{LL} this situation is characterized as a condition for the applicability of semiclassical approximation. 
Therefore, free particles in scattering experiments can be described with great accuracy in semiclassical approximation.

Let us now discuss the following question. Based on standard concepts, one might think that singletons forming a particle which is considered elementary in the standard theory, have spatial coordinates close to each other because sizes of elementary particles are considered small. As noted in Sec. \ref{statement}, at the quantum level, one can talk about physical quantities only from the point of view of operators describing these quantities. Therefore a problem remains how to interpret spatial coordinates of supersingletons that are not observed in free states. 

The same problem can be posed for baryons and mesons consisting of quarks which do not exist in free space but in the literature, position operators for such quarks are not discussed. The concept
of spacial coordinates originates from macroscopic physics and it is not clear whether this
concept still has a physical meaning for objects which do not exist in free states. 

However, the following can be noted. In scattering experiments, the coordinates of initial and final particles are large because they are of the same order of magnitude as the coordinates of macroscopic bodies. With good accuracy we can assume that the coordinates of singletons or quarks inside a bound state are approximately the same as coordinates of the entire bound state as a whole. The problem is whether the relative coordinates of singletons or quarks forming a bound state have physical meaning. As already noted, the concept of spatial coordinates arose from macroscopic physics and therefore it is not clear what physical meaning such small quantities as relative coordinates have. 
However, if we believe that the bound state consists of free singletons, then their momenta, angular momenta and coordinates are known with good accuracy. Consequently, the operators $M_{ab}$ for each singleton are known with good accuracy and these operators can be considered in semiclassical approximation.

In this approximation, the supersingleton wave functions $$\sum_{jk}c(j,k){\tilde e}(j,k)$$
are such that  the coefficients $c(j,k)$ are not equal to zero
only at $j\in (j_1,j_2),\,\, k\in (k_1,k_2)$ where $j_2-j_1\ll j_1,\,\, k_2-k_1\ll k_1$ and
the values of $|c(j,k)|$ at such $j,k$ are approximately the same. We define the angular
dependence of the coefficients as 
$c(j,k)=|c(j,k)|exp[i(j+k)\chi+i(j-k)\varphi]$.
Then taking into account Eqs. (\ref{Mab},\ref{d'},\ref{normalized}) and the definition of the basis
elements and the coefficients $c(j,k)$, direct calculation shows that, in semiclassical approximation, when the operators $M_{ab}$ can be replaced by their mean numerical values:
\begin{eqnarray}
&&L_x=2(jk)^{1/2}cos(2\varphi),\,\,L_y=-2(jk)^{1/2}sin(2\varphi),\,\,L_z=j-k\nonumber\\
&&M_{10}=jsin(2\varphi+2\chi)+ksin(2\varphi-2\chi),\,\, M_{04}=j+k+1\nonumber\\
&&M_{20}=jcos(2\varphi+2\chi)+kcos(2\varphi-2\chi)\nonumber\\
&&M_{30}=2(jk)^{1/2}sin(2\chi),\,\,M_{34}=2(jk)^{1/2}cos(2\chi)\nonumber\\
&&M_{14}=kcos(2\varphi-2\chi)-jcos(2\varphi+2\chi)\nonumber\\
&&M_{24}=jsin(2\varphi+2\chi)-ksin(2\varphi-2\chi)
\label{semiclass}
\end{eqnarray}
As can be seen from these expressions, for a single supersingleton there is no
scenario when the eigenvalues of the operators $M_{4\mu}$ are much greater than the eigenvalues of the operators $M_{\mu\nu}$ ($\mu,\nu$=0,1,2,3). This is an  
argument why singletons cannot exist in free states.

The eigenvalues of the operators in Eqs. (\ref{semiclass}) satisfy the property that when one applies the transformations 
\begin{equation}
j\leftrightarrow k,\,\,\chi\rightarrow -\chi,\,\,\varphi\rightarrow \varphi+\pi/2
\label{transform}
\end{equation}
then all the eigenvalues of the operators $M_{4\mu}$ do not change while all the eigenvalues
of the operators 
$M_{\mu\nu}$ 
change their sign.

As noted above, for a system of two free supersingletons 1 and 2, the AdS superalgebra representation is the tensor product of the representations
for supersingletons 1 and 2, and the representation operators are the sums of the corresponding operators: $M_{ab}=M_{ab}^{(1)}+M_{ab}^{(2)}$. 

If the eigenvalues of $M_{ab}^{(1)}$ are described by Eqs. (\ref{semiclass})
with the parameters $(j,k,\chi,\varphi)=(j_1,k_1,\chi_1,\varphi_1)$
and the eigenvalues of the operators $M_{ab}^{(2)}$ are described by Eqs. (\ref{semiclass})
with the parameters $(j,k,\chi,\varphi)=(j_2,k_2,\chi_2,\varphi_2)$
then, as follows from the remarks after Eq. (\ref{transform}), if 
\begin{equation}
j_2\approx k_1,\,\,k_2\approx j_1,\,\, \chi_2\approx -\chi_1,\,\,\varphi_2\approx \varphi_1
+\pi/2
\label{approx}
\end{equation}
then the eigenvalues of the operators $M_{4\mu}^{(1)}$ and $M_{4\mu}^{(2)}$ will be
approximately equal for each $\mu$ while for each $\mu,\nu$ the eigenvalues of the operators
$M_{\mu\nu}^{(1)}$ and $M_{\mu\nu}^{(2)}$ will approximately differ by sign. 
Therefore, for the operators describing the tensor product, the eigenvalues of the operators $M_{4\mu}$
will be much greater than the eigenvalues of the operators $M_{\mu\nu}$, and this
guarantees that Poincare symmetry will be a good approximate symmetry.

\section{Conclusion} 
As shown by Dyson \cite{Dyson}, it follows even from purely mathematical considerations that Poincare quantum symmetry is a special degenerate case of de Sitter quantum symmetries. As shown by Flato and Fronsdal \cite{FF}
(see also \cite{HeidenreichS}), in standard AdS theory, each massless IR can be 
constructed from the tensor product of two singleton IRs discovered by Dirac in his 
seminal paper \cite{DiracS}. As explained in Sec. \ref{supersymmetry}, AdS
theory based on supersymmetry is more general (fundamental) than standard AdS
theory.

Therefore, a question arises why in particle physics, Poincare symmetry works with a very high accuracy. The usual answer to this question is that a theory in de Sitter space becomes a theory in Minkowski one when the radius of de Sitter space becomes very large. However, de Sitter and Minkowski spaces are purely classical concepts, while in quantum theory 
(and especially in particle theory) the answer to this question must be given only in terms of quantum concepts. 

As noted in Sec. \ref{statement}, at the quantum level, Poincare symmetry is a good approximate symmetry if the eigenvalues of the operators $M_{4\mu}$ 
are much greater than the eigenvalues of the operators 
$M_{\mu\nu}$ ($\mu,\nu=0,1,2,3$).
As shown in Sec. \ref{Scenario}, for a single supersingleton there is no scenario when these conditions are met but explicit {\it mathematical} solutions with such properties exist when: 
\begin{itemize}
\item a particle which in the standard theory is considered a neutral elementary particle consists of a supersingleton and its antiparticle, and the result follows from Eqs. (\ref{AdStheta},\ref{2lambda},\ref{0}).
\item a particle which in the standard theory is considered elementary, 
consists of two supersingletons satisfying the semiclassical approximation, and 
their states satisfy the conditions (\ref{approx}). 
\end{itemize}

The title of the present Special Issue is "The Benefits That Physics Derives from the Concept of Symmetry", and the present paper shows that this concept helps solve the {\it mathematical} problem of why Poincare symmetry works with high accuracy in particle theory. There are many examples in physics when a certain problem was solved purely mathematically, but the physical meaning of the solution was understood only after some time. As noted in Sec. \ref{Scenario}, in the given problem it is not clear yet whether it is possible to physically interpret the relative position operators for singletons and quarks that are not observed in free states. 

{\bf Acknowledgments}. I am grateful to Vladimir Karmanov and the reviewers of this paper for important useful comments that were taken into account when revising the paper.

\begin{flushleft}{\bf Funding:} This research received no external funding.\end{flushleft}
\begin{flushleft}{\bf Conflicts of Interest:}  The author declares no conflicts of interest. \end{flushleft}


\begin{thebibliography}{99}
\bibitem{Novozhilov} Yu. V. Novozhilov, {\it Introduction to Elementary Particle Theory.}
International Series of Monographs in Natural Philosophy {\bf 78}, B01K2IQ5L2 Pergamon (1975). 
\bibitem{book} F.M. Lev, {\it Finite Mathematics as the Foundation of Classical Mathematics and Quantum Theory. With Application to Gravity and Particle theory}. ISBN 978-3-030-61101-9, Springer, Cham (2020). 
\bibitem{Dyson} F.G. Dyson, {\it Missed Opportunities}. Bull. Amer. Math. Soc. {\bf 78}, 635-652 (1972). 
\bibitem{Evans} N.T. Evans, {\it Discrete series for the universal covering group of the 3+2 de Sitter group}. J. Math. Phys. {\bf 8}, 170–184 (1967).
\bibitem{Lambda} F.M. Lev, {\it Solving Particle–Antiparticle and Cosmological Constant 
Problems}. Axioms {\bf 13}, 138 (2024).
\bibitem{Heidenreich} W. Heidenreich, {\it All Linear Unitary Irreducible Representations of de Sitter Supersymmetry with Positive Energy}. Phys. Lett.  {\bf B110}, 461-464 (1982).
\bibitem{Wigner} E. Wigner, {\it Normal Form of Antiunitary Operators}. J. Math. Phys. {\bf 1}, 409–412 (1960). 
\bibitem{Schwinger} J. Schwinger, {\it The Theory of Quantized Fields II}.
Phys. Rev. {\bf 91}, 713 (1953).
\bibitem{FF} M. Flato and C. Fronsdal, {\it One Massles Particle Equals two Dirac Singletons}. 
Lett. Math. Phys. {\bf 2}, 421-426 (1978).
\bibitem{HeidenreichS} W. Heidenreich, {\it Tensor Product of Positive Energy Representations of ${\tilde S}(3,2)$ and ${\tilde S}(4,2)$}. J. Math. Phys. {\bf 22}, 1566-1574 (1981).
\bibitem{DiracS}  P.A.M. Dirac, {\it A Remarkable Representation of the 3 + 2 de Sitter group}. J. Math. Phys. {\bf 4}, 901-909 (1963).
\bibitem{FFS} M. Flato, C. Fronsdal and D. Sternheimer, {\it Singleton Physics}. hep-th/9901043 (1999).
\bibitem{Bekaert} X. Bekaert, {\it Singletons and Their Maximal Symmetry Algebras}. arXiv:1111.4554 (2011).
\bibitem{Bekaert1} X. Bekaert and B. Oblak, {\it Massless Scalars and Higher-Spin BMS in Any Dimension}. arxiv:2209.02253 (2022).
\bibitem{Ponomarev1} D. Ponomarev, {\it Towards Higher-spin Holography in Flat Fpace}. 
arxiv::2210.04035 (2022).
\bibitem{Ponomarev2} D. Ponomarev, {\it Chiral Higher-spin Holography in Flat Space: the
Flato-Fronsdal Theorem and Lower-Point Functions}. arxiv:2210.04036 (2022). 
\bibitem{Bekaert2} X. Bekaert, A. Campoleoni and S. Pekar, {\it Carrollian Conformal Scalar as Flat-space Singleton}. arxiv:2211.16498 (2022).
\bibitem{supergrav} H. Samtleben and E. Sezgin, {\it Singletons in Supersymmetric Field Theories
and in Supergravity}. arxiv:2409.03000 (2024).
\bibitem{Bohm} D. Bohm, {\it Quantum Theory}. Dover Publications. ISBN 9780486659695 (1989). 
\bibitem{LL} L.D. Landau and E.M. Lifshits, {\it Quantum Mechanics: Non-Relativistic Theory}. Butterworth-Heinemann (1981).
\end{thebibliography}
\end{document}